\documentclass[aps,pre,floatfix,onecolumn,nofootinbib,showpacs]{revtex4} 
\usepackage{graphicx}\usepackage{amsmath}\usepackage{amssymb}

\newcommand{\BEQ}{\begin{eqnarray}}
\newcommand{\EEQ}{\end{eqnarray}}
\newcommand{\BEA}{\begin{eqnarray}}
\newcommand{\EEA}{\end{eqnarray}}
\begin{document}
\title{Differentiation with stratification: a principle of theoretical physics in the tradition of the memory art.}

\author{Claudia Pombo}

\address{Amsterdam, The Netherlands}



\begin{abstract}

The Art of Memory started with Aristotle's questions on memory. During its long evolution, it had important contributions from alchemist, was transformed by Ramon Llull and apparently ended with Giordano Bruno, who was considered the best known representative of this art. This tradition did not disappear, but lives in the formulations of our modern scientific theories. From its initial form as a method of keeping information via associations, it became a principle of classification and structuring of knowledge. This principle, which we here name {\it differentiation with stratification}, is a structural design behind classical mechanics. Integrating two different traditions of science in one structure, this physical theory became the modern paradigm of science. 

In this paper, we show that this principle can also be formulated as a set of questions. This is done via an analysis of theories, based on the epistemology of observational  realism. A combination of Rudolph Carnap's concept of theory as a system of observational and theoretical languages, with a criterion for separating observational languages, based on analytical psychology, shapes this epistemology. The `nuclear' role of the observational laws and the differentiations from these nucleus, reproducing the general cases of phenomena, reveals the memory art's heritage in the theories. Here in this paper we argue that this design is also present in special relativity and in quantum mechanics.

\end{abstract}

\maketitle
\section{Introduction}

Different historical accounts, based on different philosophical approaches, have been presented in the literature for the emergence of classical mechanics.
One of them assumes that the old Greek approach, supposedly based only on rational principles, could not reach the truth on nature. In this view, it was the experimentation and methods of analysis, synthesis and  induction, which led to modern science and physics is an empirical science in its essence. 
Another claim has been that it was the integration of mathematical sciences based on principles and empirical knowledge on matter, which formed the modern science. Thomas Kuhn (1922 - 1996) considered a separation between the classical and the Baconian sciences. The first tradition comprises sciences based on systematic measurements and analysis such as astronomy, statics, harmonics and optics. The old research on matter, involving observation of electric, magnetic, chemical and heat phenomena, were organized under empirical classifications and comprised the class of the Baconian sciences. Kuhn assumed that geometry was an empirical science \cite{ku1976}.
Social and political aspects have also been considered as important influence on the development of science. This idea was put forward, between others, by Edgar Zilsel (1891 - 1944) \cite{zi1942}. He argued that social pressures approximated different layers of the society such as scholars, humanists and artisans, resulting in an interchange of ideas an approaches, which characterizes the modern sciences.  

Classical mechanics emerged from an encounter between mathematized sciences and empirical traditions, but not only. It has central structure which originated from an old tradition of organization of the knowledge, known as memory art 
\cite{ro2000,ya1966,sa2013}. This discipline originated in the Greek analysis on natural memory  
and developed through the times, resulting in a structural model for knowledge resembling the alchemical structural model of matter. 
The main new contribution in this paper is to show this connection by recovering the method of the memory art in explicit terms. 
In order achieve this objective, we analyze the science which came later, by means of a suitable epistemology, named observational realism 
\cite{po2010,po2011}.

As discussed in these references, this epistemology integrates the theoretical model proposed by Rudolf Carnap (1891 – 1970) with the psychic structure conceived by Carl Gustav Jung (1875 – 1961). The first distinguished two different sectors in theories: observational and theoretical. We then justify this differentiation by identifying the observational sector as the manifestation of an archetypal or innate form of language. 
The possibility of innate ideas in physics was put forward by Jung and Wolfgang Pauli (1900 – 1958) in collaboration \cite{me2001,atfu2014}. 
Carnap's separation of the observational and theoretical domains, via corresponding concepts and languages, is clearly reflected in the set of laws of classical mechanics. And the interrelation of the laws is built according to the principle of `stratification with differentiation', which is the explicit name for the last stage of the memory art. We also discuss aspects on the foundations of two other theories of physics, special relativity and quantum mechanics, showing their structural similarity.

\section{Origins and evolution of the memory art.}
\label{Memory art}

Memory nowadays is a concept belonging to all branches of science, from psychology and sociology, to computer science, biology and physics. Still, it is a difficult notion to approach and define. On the other hand, natural memory usually means the ability, found in animals and humans, to recall experiences and knowledge. This phenomenon called the attention from early philosophers who also had interest to find methods to improve its performance. While at present times one can buy and carry small and light books, computers and cds, to access knowledge for professional and normal daily purposes, these technologies were absent in the past. Memory started to be explored and developed as a support discipline, with broad application in all professions and intellectual and fields. 

Aristotle dedicated a complete work to this topic. He suggested that memory should be more than recollection of experiences \cite{ar2011}, suspecting that its understanding could be a key for prediction. And from Aristotle's reflections, emerged the art of memory, an intellectual tradition of improvement of memory, which passed through transformations. As a tool for organization of the knowledge, it was later influenced by alchemists. Between them, Augustine of Hippo (354-430), Albertus Magnus (1193 or 1206-1280), Thomas Aquinas (1225-1274), Ramon Llull (1232-1315) and Giordano Bruno(1548-1600). During its development, this art absorbed the alchemical paradigm of matter, a stratified model of material structure, in its imagery. Llull not only introduced this aspect in it, but he explored this change, making from it a basis for the Llullian art. After this, Bruno became known as one of the greatest and last practitioners of the art of memory \cite{ya1964}, passing this knowledge on to Isaac Newton and G. Leibniz. 

But the practice of the art of memory was not explicitly summarized by Bruno and this led to the belief that it vanished. However, considering the discovery of the differential calculus, independently by both Newton and Leibniz and the design of the theory of classical mechanics by Newton and other scientific advances of the modern times, we can reconstruct the principles of this art. Moreover, not only in Llull's works one can find  the foundations of Bruno's practices but his combinatorial art had great influence on Leibniz, who contributed to the development of the science of logic. The influence of Llullism on the emergence of the abstract algebra which took place in the end of the nineteen century via Leibniz's, is now a topic of interest in recent literature \cite{fisi2011}. Apart from this, Leibniz was greatly influenced by Aristotle's ideas on conservation, potentiality and actuality, opening a new approach for the physics of matter, not treated by Newton's physics. 

The principle of differentiation with stratification is a description in words of Llull pictorial messages, it is a formulation of the late stage of the memory art. It proposes a nucleus, differentiating from other elements, similar to the paradigm of the unification of the matter in the tradition of alchemy. The evidence that psychology and alchemy had possibly common principles, was realized by Jung, being a key element for the recovery of this art, by comparing structure of ideas and structure of matter. The art is presented by means of a pair of questions: \\
$(1)$- What is the common or essential aspect in a diversity of considered objects or phenomena?\\
$(2)$- How to organize the differences between them?

The first question specifies the class of objects or phenomena under treatment by the method. The answer to this question reveals an ideal form,
different from all of the others, but at the same time in them all. In this sense, it stratifies the object or phenomenological domain, by detaching and centralizing this class. The second question asks for the expression of the deviation from the central one. 
 
In the context of physics, we will identify these answers as the observational and the theoretical laws of a theory, as considered
by Carnap. The answers to the first and second questions are not unique. Different theories apply to different classes of phenomena.
In this epistemology, a theory is complete if the causes of the differentiations are explicitly given and the phenomena are reproduced. 

\section{Physics and its early origins and foundations.}
\label{Physics, origins and foundations.}

In one or another form, space has been present as an element in all cosmogonies of creation since the antiquity. Space does not emerge alone but together with its partner matter. We assume that they comprise a pair of archetypal patterns. As a concept, it is well known that space is not only one and the same holds for matter. We assume that physical references comprise a number of archetypal pairs, such as change-permanence, continuity-discreteness, between others. However, here in this paper we will focus on the representations of the space-matter pair only.

In the domain of the physical sciences, different concepts of space have been formulated as history developed. These concepts are introduced by means of the formulation of processes or forms of animation of the matter.
In this paper, we use the word animation for variations without active influences. Movement is only one between other forms of animation.
In history, a metrical space was naturally introduced in connection with movement, the latter was the first phenomenon to acquire a complete theoretical treatment and this led to the belief that physics developed either by generalizing its phenomenological domain, or by extending its theoretical treatment.
Physics comprises different theories with a common structural design, but having separate observational domains, due to differences in their models of space and matter. We can turn to history to show that differences were established very early in the history of knowledge.

Philosophers in the past argued that, behind the apparent diversity of the observed matter, there should be a sort of permanent essence in it. This naturally led to the question of how, from this hypothetical essence, all the matters could come about. Observing processes of transformations of matter, some philosophers then proposed that fire, water, air and solid matter were essential aspects of matter and named them the elements. These last philosophers were the precursors of the alchemist, for whom there should also be an unity in the material world a part from the elements. This idea became a principle in alchemy and this unity was named {\it prima materia}. Another ancient school of philosophers, which came to be known as atomist and grounders of the philosophical materialism, claimed that, in essence, moving corpuscules should be all in the universe. These ideas were also criticized by not being able to explain the diversity of the material forms. According to some philosophers, a certain diversity and structure should be added to the corpuscles. Apart from these schools, there were also philosophers who claimed that numbers were the final essence of the existences. For these thinkers, the borders between the material world and the world of ideas was not so sharp as for the previous ones. The connection between material phenomena and ideas was scrutinized Plato, who proposed classifications in the world of the ideas, to approach the interface of these worlds. 

During these times, knowledge of geometry was in development by empirical means. Then, Plato (427— 347 BC) conceived the unification between separation and extension of bodies. This is the reason why Aristotle (384–322 BC) said that, among all the philosophers who considered the question of space, only Plato had actually understood it \cite{ar1999}. But Aristotle himself was the on who opened the field of natural philosophy, suggesting a mathematization of material processes, in order to express conservations. He made a careful analysis on the philosophical claims of his predecessors, preoccupied with the establishment of a sharp separation between the natural and the non-natural world. For this, he discussed an important difference between the notions of change and variation, by emphasizing the presence of a subject in the latter. He and his successors had many questions about material models which could describe and treat material phenomena, also focusing on the question of the movement. With respect to movement, its intrinsic relativity was not understood at that time. The rectilinear uniform movement was not formulated or assumed to be a natural principle, in spite of the fact that a mathematical treatment for it, in the form of proportion, did emerge. 

Soon later, Euclid formalized the science of geometry by treating Plato's unification under Aristotle's logic. The Euclidean geometry became a foundation for the natural sciences. Among the classical sciences, there was geometrical optics, considered the science of the visual rays. A physical nature for light was searched but no consensus was reached on this subject. Light was though to be a bunch of corpuscles, also ethereal matter, while ether was also claimed to be non-material. An important part of the scientific program, concerning the mathematization of matter and its processes was developed by Archimedes who adopted an empirical approach to geometry and mathematics. 
Parallel to these developments, alchemy started its development as the science of matter, being connected to a broad range of the natural sciences, from electrical, magnetic and luminous phenomena to physiology and medical sciences. Observations and experiments connecting luminous phenomena and heat led to questions on the nature of celestial objects. But alchemy was detached from mathematics, mainly because its concept of space of influence had not a geometrical treatment. 

For some reason, those natural philosophers and mathematicians from the past, who developed the sciences of numbers, geometry and the corresponding algebras, did not pay too much attention to an useful form of thinking which was already known in their time \cite{bo1968}. 
This was our familiar table representation, for the purpose of comparison and analysis. Long before appearing in Greece, it was used and studied in China. Originally, a table consists of associations of classes versus conditions. Being mathematized, by having quantified these associations, tables were mostly used in accountancy activities. In Greece, they were not considered as mathematical concepts on their own, and this view on them was kept till the end of the 19th century. Then, a special algebraic domain was established for them, consisting of operations and applications. This provoked a complete review on the concepts of algebra and also of logic.
When the mathematization of the natural processes was suggested by Aristotle, the mathematics in question did not include a matricial algebra which could serve as a basis of representation of discrete forms of matter and their processes. This representation took a long time to be introduced but its basis had been always present. In this sense, one can say that our modern atomic theory has very old roots, as old as the roots of the science of the movement can be.

\section {Space versus matter in classical mechanics, in special relativity and in quantum mechanics.}
\label{space versus matter}

Reviews on Aristotle's treatment of movement were done later, specially by John Philoponus (490 AC– 570 AC), who introduced an empirical treatment to it. In the XIV century in Oxford emerged the Merton's Rule, involving averages of velocities with constant acceleration. In France, Nicole Oresme (1323-1382), reviewed an argument of Aristotle on the quantification of variable forms and continuity. From these works an average velocity was defined, leading to the definition of constant velocity in uniform movement. But the consequences of these results for physics, implying the absence of material influences on this kind of movement, was not recognized. Oresme held that all which can be measured is continuous \cite{bo1968}. 
Finally, around Galileo's times, the relativity of the velocity, that means its independence from material influences, was established. It then became a physical principle in Newton's theory. At the same time, Leibniz and Newton established the differential calculus \cite{bo1959} by means of the memory art. Considering the straight line as the nucleus from which the class of all curved lines is deviated, the concept of derivative is a measure of the differences, according to the art. It is very interesting to notice that Leibniz used the name differentiation for this analytical process.

Newton's velocity integrates the Platonic unified length ($l$) with material cycles. The interpretation of these cycles as a measure of absolute time ($t$) leads to a concept of physical space. The usual geometrical approach of time, meaning its representation by a straight line, is granted by this relation. Newton's theory goes further, by introducing the alchemical space of influence to physics. This was done by connecting empirical forms of 
force with the force of inertia. The third law is a statement of conservation and it rules the empirical sector of the second law. The expression {\it minimum animation} fits well for a nuclear concept of movement. Then, for the first question
{\it `Which is the common feature which characterizes all movements?'}, 
the answer is {\it `To be deviations of the rectilinear uniform movement'.} 
And to the second question 
{\it `How to establish their differences?'}, 
the answer is {\it `By specifying the causes of the deviations'}. 
These causes are forces of interaction, empirical constructs which can be found via experimentation. In equations we can write the observational law as
\BEQ d=v t. \EEQ 
 
The case of special relativity can be considered as a modification of the previous observational sector. To the distance formed by movement is associated a concept of formation of signals, resulting in a new form of geometrical space. In order to compare these distances we recall that they originate from unification and integration of very different concepts of space: material extensions, separations, paths and formation of signals. The relativistic distance can be written in this form below:
\BEQ D=(c^2 {t'}^{2}+v^2 t^2)^{1/2} \EEQ
for $D=ct$. This makes explicit the double reference of time in the relativistic distance $D$, due to the introduction of a form of observation which was absent in the previous case \cite{poni2006}. In this reference, the authors explained how $ct'$ is observed by means of local detection and that this fact does not imply a connection to perception. The distance $ct'$ is out of Plato's unification, resulting that $D$ is not a distance measurable by means of material extensions. In a limit, the relativistic concepts of distance equals the classical distance.

In the classical theory, continuity of matter and space are related. If space of separation is continuous, its divisibility has no limit other than to reach the material realm of extension and vice versa. If extension is not continuous or divisible, the space of separations contains holes of material stuff and therefore it would not be continuous either. In classical physics, matter can appear in observational and empirical forms such as extension, shape, inertial mass and interactive substance, such as gravitational mass and electrical charge. With the unification of the spaces of movement and interaction, all these forms of matter became continuous. In which concerns to the relativistic case, because the classical limit is part of it, these conditions also hold, taking apart the possibility of non continuous signals, which we do not discuss here. 

For approaching quantum mechanics, we must pose the question of which discreteness is introduced by this physics, if this is done in the observational or in the theoretical realm. Discussions on the subject of the quantum reality had not clarified the nature of this quantization, if observational or in the domain of the causes. The quest for the quantum reality in connection with the quantum language, as posed in the influential EPR paper \cite {epr1935} and many others which followed, did not make this distinction. Questions on measurement, determinism and causality need this separation. 

Then, before to enter in these questions, we firstly argue that quantum mechanics did not emerge from the classical sciences which led to the theory of the movement, but developed from a branch of the sciences of matter, which had its own investigation. This science was lacking a mathematical treatment and its descriptions of matter did not involve a concept of material extension. To enter in physics, or to connect with the Platonic geometrical world, it had to give a geometrical or a mathematical form to matter. Inspired by the principle of differentiation with stratification, we conjecture that its main question was to model the reproduction of material forms, from a non-material nucleus. This was the model of thinking, apart from specifying the nature of this nucleus and of these matters. 

We can follow some clarifying arguments presented by Paul Dirac, one of the main developers of quantum mechanics \cite{di1978}. He said that in 1925 Heisenberg introduced a matricial form for positions, distances, velocities, momentum and other physical variables. In fact, after Leibniz, matrices had been used in physics, in connection with algebraic and vectorial representations. But a position is known to belong to a space of separation and, in the Platonic space, it was geometrized with the extensions. A position has sense as a line either in a three-dimensional or in a n-dimensional space. So, we only can find sense in Heisenberg construction, if it is to hold before Plato's unification, in the strict domain of the extensions. This is the point which we would like to highlight here. As far as we know, Plato, Aristotle, Euclid and the other mathematicians from the past, did not question the possible existence of an internal reality for material extensions. When Heisenberg introduced this physics, he also did not explain how this should be understood. Taking for granted his construction and re-integrating it with Plato's unification, a separation $s$ becomes the result of a re-interpretation of the matricial extension $E$. This re-interpretation means the introduction of $E$ in a geometrical space, as below:  
\BEQ \label{3}
s=
\left(\begin{array}{cccc} m_1 & m_2 & . & . \end{array}\right)\cdot
\left(\begin{array}{cccc} e_{11} & e_{12} & . & . \\ e_{21} & e_{22} & . & . \\ . & . & . & . \\ . & . & . & . \end{array}\right)\cdot
\left
(\begin{array}{c} n_1 \\ n_2 \\ . \\ . \end{array} \right)
\EEQ

After this, we can compare the three forms of distance ($d$, $D$ and $s$) in their respective equations (1),(2) and (3), introduced by each theory, as a first step in the comparison of their observational domains. Apart from the possible meanings of the extra vectorial elements, introduced as mediators of the geometrization process in equation (3), we can see how special relativity and quantum mechanics diverge in their spatial settlements. The geometrical basis of special relativity approaches Euclidean geometry from the outside while the quantum basis approaches from the inside. 

A matrix is a space of associations. It is considered to be intrinsically connected with physical discreteness. In the mathematical construction of equation (3), the matrix $E$ plays a central role in the formulation of extensions. It can have any number of dimensions, provided that its accompanying vectors share the same number. Here we have a space of associations in a nuclear position, in the construction of material extensions and physical distances. Then, the matrix, meaning the matricial representation, is the observational nucleus which we were looking for. In our epistemology we do consider observation much beyond the usual concept of measurement. But still it is puzzling to realize that an element which stands for the organization of thinking can also be an explicit stem of material extension and therefore of physical phenomena. 

\section{Final comments.}

We cannot give here a proper answer to the second question posed by the memory art, but only propose to search for it. 
The principle of {\it differentiation with stratification} has given structure to physical theories and we expect that it will also apply to quantum mechanics. 

The main objective of our program is to detach and compare the observational domains of physical theories, in order to study its process of evolution. Concerning classical mechanics and special relativity, this discussion has been presented in a series of papers. Now we present a few arguments to initiate an approach to the foundations of quantum mechanics by the same method. But concerning this concept of evolution here mentioned, we must be careful with its meaning, which we adopt in an epistemological sense only. History and foundations of science are subjects apart. Considering quantum mechanics, it would be reasonable to say that it has pre-Platonic roots, in spite of the fact that it has been characterized as a XX century materialistic science.

\begin{acknowledgements}

The author would like to thank Xander Giphart, Isabela Pombo-Geertsma and David Costa for discussions on the topics developed here. 
\end{acknowledgements}

\end{document}